\documentclass[%
    aps, 
    prab, 
    twocolumn, 
    superscriptaddress, 
    longbibliography,
    floatfix,
    10pt
]{revtex4-2}
\usepackage[colorlinks=true, linkcolor=blue, citecolor=blue, urlcolor=blue]{hyperref} 
\usepackage{xurl} 
\usepackage{graphicx}
\usepackage{tikz}
\usepackage{subcaption}
\usepackage{dcolumn}
\usepackage{bm}
\usepackage{amsfonts,amssymb}
\usepackage{overpic}
\usepackage{relsize}
\usepackage{caption}
\usepackage{booktabs} 
\usepackage{array}

\begin{document}


\title{Neural network-based deconvolution for GeV-Scale Gamma-Ray Spectroscopy}

\author{Zhuofan Zhang} 
\author{Mingxuan Wei} 
\affiliation{State Key Laboratory of Dark Matter Physics, Key Laboratory of Laser Plasma (MoE), School of Physics and Astronomy, Shanghai Jiao Tong University, Shanghai 200240, China}
\author{Kyle Fleck}
\affiliation{School of Mathematics and Physics, Queen’s University Belfast, BT7 1NN Belfast, United Kingdom}
\author{Jun Liu} 
\email{liujun@nint.ac.cn}
\author{Xinjian Tan} 
\affiliation{State Key Laboratory of Intense Pulsed Radiation Simulation and Effect, Northwest Institute of Nuclear Technology, Xi’an 710024, China}
\author{Gianluca Sarri}
\affiliation{School of Mathematics and Physics, Queen’s University Belfast, BT7 1NN Belfast, United Kingdom}
\author{Wenchao Yan} 
\email{wenchaoyan@sjtu.edu.cn}
\affiliation{State Key Laboratory of Dark Matter Physics, Key Laboratory of Laser Plasma (MoE), School of Physics and Astronomy, Shanghai Jiao Tong University, Shanghai 200240, China}
\affiliation{Collaborative Innovation Center of IFSA, Shanghai Jiao Tong University, Shanghai 200240, China}

\begin{abstract}
High-energy gamma-ray spectroscopy is crucial for studying and advancing the application of high-energy photons in areas like strong-field physics, high-energy-density science, and laboratory astrophysics. However, high-energy gamma-ray spectroscopy in the multi-MeV to GeV range faces significant challenges in precise spectral reconstruction. This study presents a machine learning-based inversion approach that combines a spectrometer design with advanced deconvolution algorithms. We develop a gamma-ray spectrometer optimized through Monte Carlo simulations for maximum positron yield and minimal noise. A two-stage neural network framework is proposed based on the structure of the spectrometer: a denoising autoencoder suppresses statistical noise in measured positron spectra, while a U-Net architecture solves the ill-posed inverse problem to reconstruct incident gamma spectra. This approach establishes a new methodology for gamma-ray diagnostics in strong-field QED experiments and high-energy photon sources.
\end{abstract}

\maketitle

\section{introduction}\label{introduction}
High-energy gamma-ray spectroscopy in the multi-MeV to GeV range is essential for advancing our understanding of fundamental physics and emerging applications, from astrophysics to strong-field quantum electrodynamics (SFQED). Despite significant progress in gamma-ray source development, such as bremsstrahlung emission from laser-wakefield accelerated electron beams \cite{liBetatronRadiationBremsstrahlung2023,lemosBremsstrahlungHardXray,nohChargeneutralGeVscaleElectronpositron2024,lemosUltrabroadbandXraySource2024} and inverse Compton scattering \cite{sarriUltrahighBrillianceMultiMeVgRay2014,yanHighorderMultiphotonThomson2017,zhou2023gamma,mirzaieAllopticalNonlinearCompton2024a}, precise spectral reconstruction of high-flux gamma-ray beams remains a challenge. The filter-stacked spectrometer consists of multiple filter layers and a recording medium, commonly used for detecting hard X-rays and gamma rays in the range of several to tens MeV \cite{nolte1999tld,behrens2002tld,chen2008bremsstrahlung,behrens2009spectrometer}. However, its spectral resolution is very low in the tens of MeV to even GeV energy range. Recent studies have demonstrated \cite{schumaker2014measurements,barbosa2015pair,singhCompactHighEnergy2024} that gamma-ray spectrometers can be developed by analyzing secondary particle energy spectra resulting from gamma-ray interactions with a converter medium.

For instance, Compton scattering in low-Z materials has been used to convert gamma-ray energy into measurable electron spectra \cite{corvanDesignCompactSpectrometer2014,zhangCompactBroadbandHighresolution2021,hadenHighEnergyXray2020,kojimaDevelopmentComptonXray2016,naranjoComptonSpectrometerFACETII2021}, enabling gamma-ray detection in the 0.5 MeV to 30 MeV range with compact magnetic systems and detectors. At higher energies,  pair spectrometers \cite{fleckConceptualDesignHighflux,cavanaghExperimentalCharacterizationSingleshot2023,abramowiczTechnicalDesignReport,songCompactLownoiseGeVscale2025} based on the pair production process in high-Z materials (e.g., tungsten, lead) have been proposed. These spectrometers convert gamma-rays into electron-positron pairs, with the original spectra reconstructed by analyzing the energy and momentum distributions of the pairs with a magnetic spectrometer. The primary focus is on the precise measurement of high-energy gamma-rays, reaching GeV or even 10 GeV levels. The converter thickness is crucial for the accurate measurement of the energy spectrum. Techniques using thin converters to suppress multiple scattering offer advantages in high-flux scenarios \cite{abramowiczTechnicalDesignReport}. However, many experiments still require thicker converters during operation because they are advantageous or indispensable under certain conditions. In unknown-flux or single-shot experiments, thicker converters are often employed to increase the reaction probability, leading to higher positron or electron yields, despite introducing more complex energy degradation. The choice of converter thickness thus involves a trade-off between resolution and yield, depending on the photon flux and the specific diagnostic requirements \cite{corvanDesignCompactSpectrometer2014}. This has created a demand for spectrometer design and reconstruction algorithms.

Several traditional algorithms, including iterative reconstruction \cite{kimComprehensiveReviewCompton2024}, Tikhonov regularization \cite{zhangCompactBroadbandHighresolution2021}, and maximum likelihood estimation \cite{lowellMaximumLikelihoodCompton2017}, are effective for Compton spectrometers. However, they provide only approximate solutions for the spectra of GeV gamma rays. Statistical Bayesian approach \cite{dagostiniImprovedIterativeBayesian2010a,cavanaghExperimentalCharacterizationSingleshot2023,valsdottirExploringFullyBayesian} and Machine learning methods like fully connected networks \cite{yadavReconstructingGammarayEnergy2024} have been explored for reconstruction in the GeV range. 
It is envisioned that the statistical Bayesian algorithm can be further improved by combining the Bayesian nature of this method with machine learning techniques \cite{fleck2025strong}. Fully connected neural networks, with their simple architectures, struggle to capture complex features such as nonlinear Compton scattering and LWFA-driven strong bremsstrahlung gamma-rays in the presence of high noise, leaving room for improvement.

Here, we present a novel approach for gamma-ray spectroscopy that combines a recently proposed spectrometer design with a corresponding machine learning approach for extracting photon energy spectra. Building upon the geometric structure of gamma-ray spectrometers, we frame spectral reconstruction as a high-dimensional inverse problem, developing a data-driven framework that outperforms traditional regularization techniques in both accuracy and applicability. Using Monte Carlo simulations, we quantify the energy-dependent response of the spectrometer system to incident photons. We introduce an end-to-end machine learning network architecture, consisting of a denoising autoencoder and a reconstruction U-Net, to effectively separate overlapping cascade features while preserving physical consistency.

The paper is organized as follows: Section \ref{OPTIMIZATION OF THE SPECTROMETER} outlines the spectrometer design and parameter optimization. The gamma-ray energy spectrum extraction procedure is described in Section \ref{Learning procedure of gamma spectrum deconvolution}. The comparison and evaluation of reconstruction results are presented in Section \ref{evaluation results}. Finally, Section \ref{Conclusion and Outlook} concludes the study and provides an outlook on future developments.
\section{DESIGN OF THE SPECTROMETER}\label{OPTIMIZATION OF THE SPECTROMETER}
\begin{figure}[htbp]
    \centering
    \includegraphics[width=1.0\linewidth]{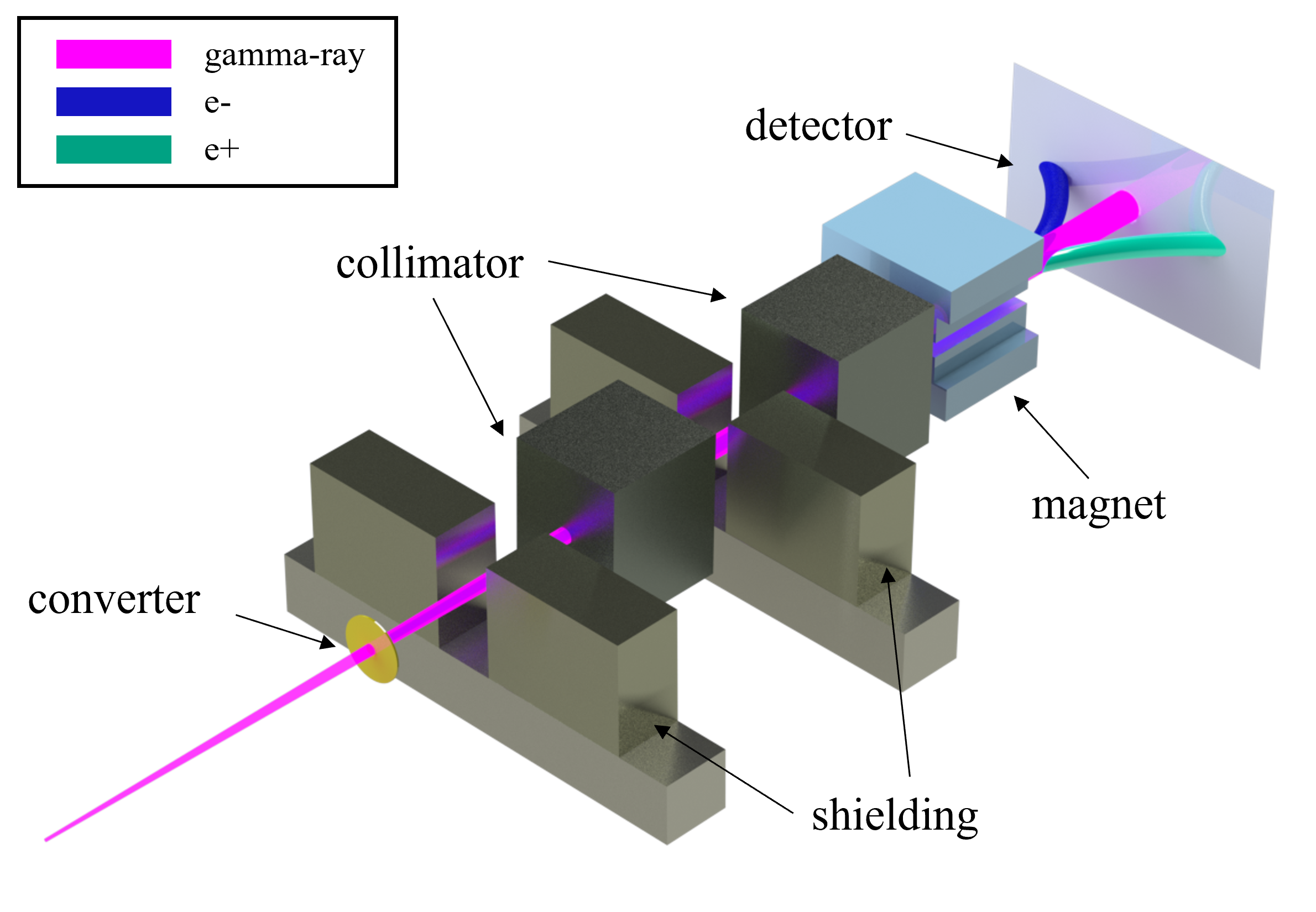}
    \caption{Top view schematic of the gamma-ray spectrometer, consisting of the converter, shielding, collimator, magnet, and detector.}
    \label{fig:1}
\end{figure}
The overall mechanical structure of the spectrometer is shown in Fig.\ref{fig:1}. The spectrometer consists of three components: the converter, the collimator, and the pair magnet spectrometer. The converter converts a small percentage of the incident high-energy photons into electron-positron pairs \cite{fleckConceptualDesignHighflux}, and the collimating aperture shields the system from major noise (e.g., off-axis photons and low-energy electrons). The electron-positron pairs are collected by detectors symmetrically distributed around the magnetic spectrometer. The double collimator configuration has proven to be ideal \cite{fleckConceptualDesignHighflux,songCompactLownoiseGeVscale2025} for minimizing the background on the detector when measuring GeV-scale electron–positron pairs. 

\begin{table}[!htbp]
\centering
\tabcolsep=0.6cm
\caption{Atomic numbers and a normalized parameter ($Z^2\rho/M $) for different target materials. This parameter is proportional to the theoretical positron yield per incident gamma-ray photon. The parameters are normalized with the value of gold as the maximum value for the sake of comparison.}\label{tab:tableTab}
\scalebox{1.0}{
\begin{tabular}{l c c c}
\toprule\toprule
\textbf{Element}    & \textbf{$Z $}      & \textbf{$Z^2\rho/M $}  \\
\midrule
Gold (Au)         & 79      & 1.0              \\
Tungsten (W)      & 74    & 0.94                  \\
Tantalum (Ta)     & 73      & 0.8             \\
Lead (Pb)         & 82   & 0.6                     \\
\bottomrule\bottomrule
\end{tabular}}
\end{table}

\begin{figure*}[htbp]
    \centering
    \begin{subfigure}{0.5\textwidth}
        \centering
        \begin{tikzpicture}
                \node[anchor=south west,inner sep=0] (image) at (0,0) {\includegraphics[width=\linewidth]{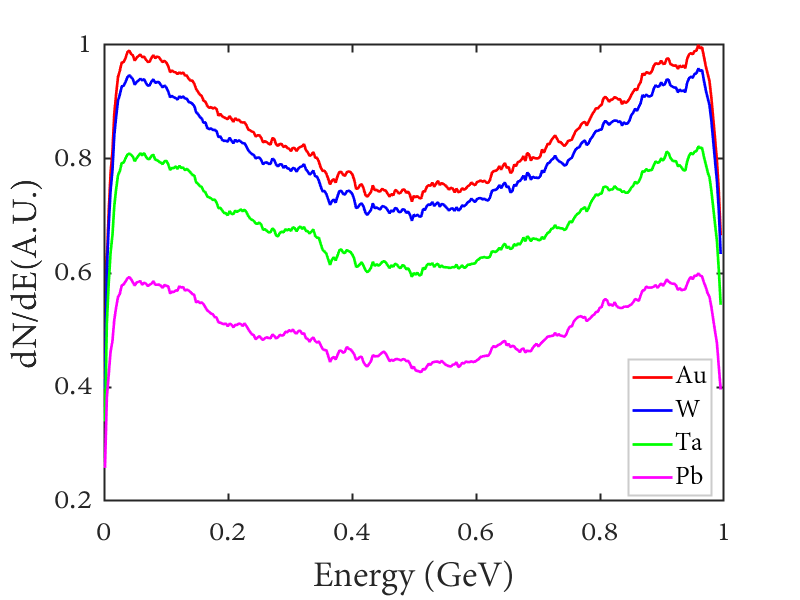}};
                \node[anchor=north west, font=\large\bfseries, yshift=0.3cm] at (image.north west) {(a)};
        \end{tikzpicture}
        \caption*{} 
        \refstepcounter{subfigure}\label{fig:elespec}
    \end{subfigure}\hfill 
    \begin{subfigure}{0.50\textwidth}
        \centering  
            \begin{tikzpicture}
            \node[anchor=south west,inner sep=0] (image) at (0,0) {\includegraphics[width=\linewidth]{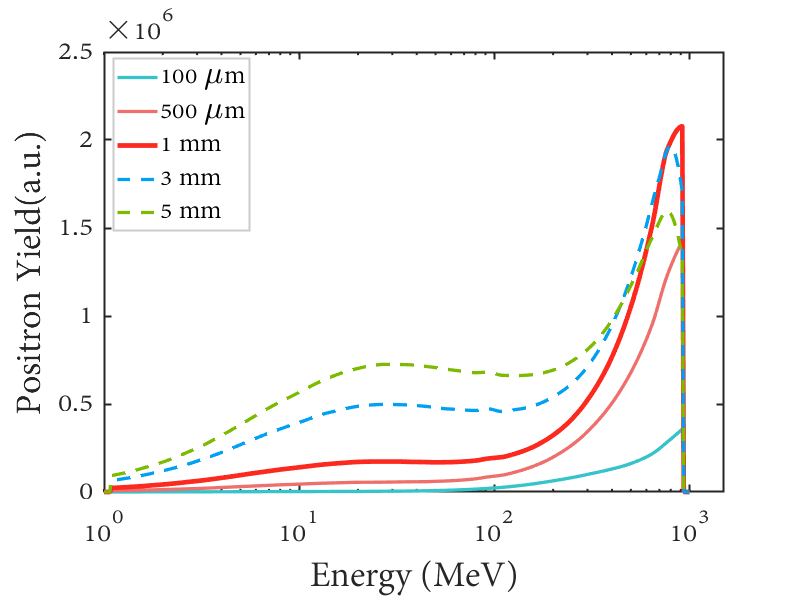}};
            \node[anchor=north west, font=\large\bfseries, yshift=0.3cm] at (image.north west) {(b)};
        \end{tikzpicture}
        \caption*{} 
        \refstepcounter{subfigure}\label{fig:yield}
    \end{subfigure}
    \vspace{-1cm}
    \caption{(a) The normalized spectra of positrons produced by a 1 GeV monochromatic gamma-ray beam after propagating through different 500 $\mu$m thick materials; (b) Spectra of positrons generated when a pencil-shaped monochromatic gamma-ray beam of energy 1 GeV penetrates tungsten targets of thicknesses 100 $\mu$m, 500 $\mu$m, 1 mm, 3 mm, and 5 mm. The x-axis represents the incident photon energy (in logarithmic scale), while the y-axis shows the number of positrons produced per bin of incident gamma-ray photon energy. As the tungsten layer thickness increases, the full width at half maximum (FWHM) of the peak decreases to 1 mm point, and then starts to increase.}
    \label{fig:2}
\end{figure*}
The gamma-ray spectrometer is based on the process of pair production in a nuclear field. Neglecting the recoil energy of the nucleus, an electron-positron pair may be produced in the Coulomb field of the nucleus only after the photon energy exceeds the rest mass of the pair by $2m_ec^2$, where $m_e$ is the rest mass of the electron and $c$ is the speed of light. As photon energy increases, pair production begins to dominate photon-matter interactions above approximately 50 MeV, overtaking Compton scattering as the primary process in high-Z materials \cite{corvanDesignCompactSpectrometer2014}. In the ultra-relativistic approximation, the total cross section \cite{berestetskiiQuantumElectrodynamicsVolume2012,fleckConceptualDesignHighflux} for pair production in the nuclear field can be expressed as 
\begin{equation}
    \sigma\approx\frac{28}{9}\alpha Z^2r_e^2\left[\log\left(\frac{2E_\gamma}{m_ec^2}\right)-\frac{109}{42}\right]\label{eq:2}
\end{equation}
where $\alpha \approx$  1/137 is the fine structure constant, $r_e$ is the classical electron radius, $Z$ is the atomic number of the converter material. 

Due to additional processes in the converter that may cause electron generation or scattering, the electron spectrum is typically affected by higher noise levels on the detector plane \cite{cavanaghExperimentalCharacterizationSingleshot2023}. Therefore, it is preferable to pay attention to the positrons and maximize their yield. The yield of high-energy positrons generated via Bethe-Heitler process is proportional to the product of the pair production cross section and the number density of the target nucleus $N_A\rho/M$, where $N_A$ is Avogadro's constant, $\rho$ is the material density, and $M$ is the molar mass. Eq.\ref{eq:2} shows that the pair production cross section is proportional to the square of the atomic number of the target nucleus. Therefore, to maximize the probability of pair production and consequently the yield of measurable positrons, it is necessary to select materials with high atomic numbers as target materials. Typically, high-Z elements suitable for target materials include gold, tungsten, tantalum, and lead. According to theoretical calculations, the gold target had the highest $Z^2N_A\rho/M $ value among the four metal targets mentioned before. Thus, gold produced the highest positron yield, followed by tungsten. To study the energy spectrum of positrons after a 1 GeV monochromatic gamma-ray beam passes through different 500 $\mu$m thick materials, simulations are performed using the Monte Carlo (MC) particle tracking code Geant4 \cite{agostinelliGeant4aSimulationToolkit2003}. Our simulation utilizes the physics package \texttt{QGSP\_BERT} \cite{CharacterizationRelativisticElectron}, which incorporates energy loss processes for electrons and positrons, including bremsstrahlung, Coulomb scattering, and ionization. We assume pencil-shaped photon beams with no initial divergence in our simulations. The Tab.\ref{tab:tableTab} and Fig.\ref{fig:elespec} show the $Z^2\rho/M $ value and the normalized positron energy spectra produced by a 1 GeV monochromatic gamma-ray beam propagating through 500 $\mu$m thick layers of different target materials. The simulation results are consistent with the numerical estimates. We chose tungsten as the converter material for its sustainability compared to gold.

The relationship between converter thickness and yield \cite{wengDesignRadiationConversion2021} is a key factor in optimizing spectrometers. MC simulations can be used to analyze the relationship between the total positron yield and tungsten target thickness under different energy pencil-shaped gamma-ray incidence conditions. In our simulation, a $10^8$ pencil-shaped monochromatic gamma ray beam with an energy of 1 GeV was simulated penetrating tungsten targets of thicknesses 100 $\mu$m, 500 $\mu$m, 1 mm, 3 mm, and 5 mm. A divergent gamma-ray beam will increase the divergence of the electron-positron pairs and thus reduce the spectral resolution of the system. We add a 5 mm collimated aperture behind the target to minimize the secondary-effect tails of tens of MeV. Fig.\ref{fig:yield} plots the positron yield as a function of incident gamma photon energy. A key parameter to define in this figure is the size of the energy bin corresponding to the number of positrons. It is not meaningful to select an energy bin size smaller than the spectrometer's intrinsic energy resolution(See Eq.\ref{resolution}). The advantage of this binning method is that it allows the optimal converter thickness to be selected based on the full width at half maximum (FWHM) of the positron yield distribution. As shown in the figure, the spectral broadening arises largely from the scattering of positrons as they pass through the converter. Hence, there is a compromise between spectral resolution and yield. As the tungsten layer thickness increases, the FWHM of the energy spectrum gradually broadens. However, the spectral peak in the high-energy region does not present a single increasing trend; instead, an optimal peak exists. It is worth noting that the radiation length of tungsten is 3.5 mm \cite{nisttungsten_W}. For millimeter-thick target materials, the thickness is comparable to the radiation length, requiring careful selection of thicknesses exceeding the radiation length. The thickness of tungsten depends largely on the photon flux to be investigated. 1 mm tungsten converter thickness provides the optimal balance between positron yield and energy resolution, minimizing spectral broadening while ensuring sufficient signal intensity. For our application, a compromise thickness of 1 mm can be selected.
\begin{figure*}[htbp]
    \centering
    \begin{subfigure}{0.5\textwidth} 
        \centering
        \begin{tikzpicture}
                \node[anchor=south west,inner sep=0] (image) at (0,0) {\includegraphics[width=\linewidth]{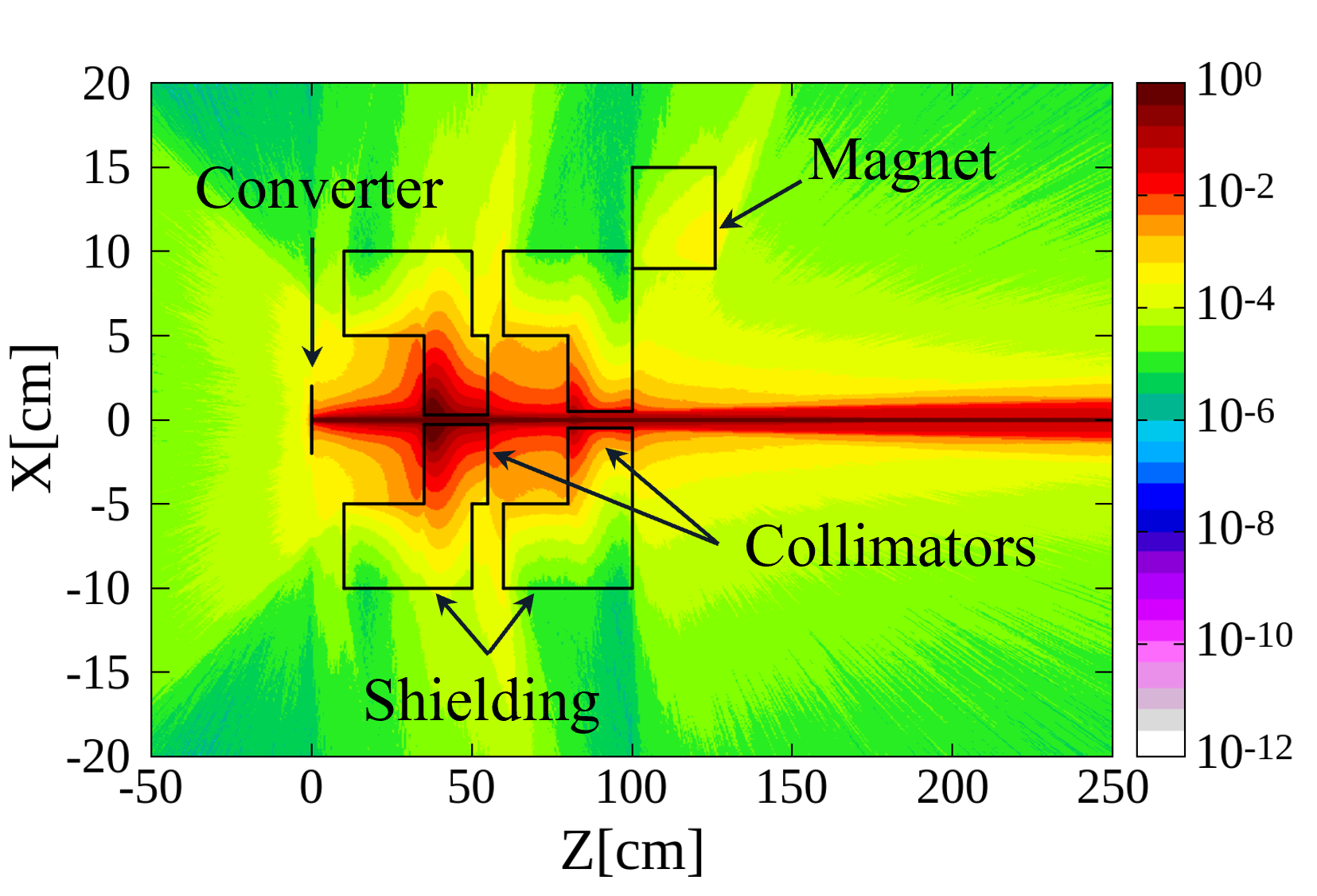}};
                \node[anchor=north west, font=\large\bfseries, yshift=0.3cm] at (image.north west) {(a)};
        \end{tikzpicture}
        \caption*{} 
        \refstepcounter{subfigure}\label{fig:phodis}
    \end{subfigure}\hfill 
    \begin{subfigure}{0.5\textwidth}
        \centering
        \begin{tikzpicture}
                \node[anchor=south west,inner sep=0] (image) at (0,0) {\includegraphics[width=\linewidth]{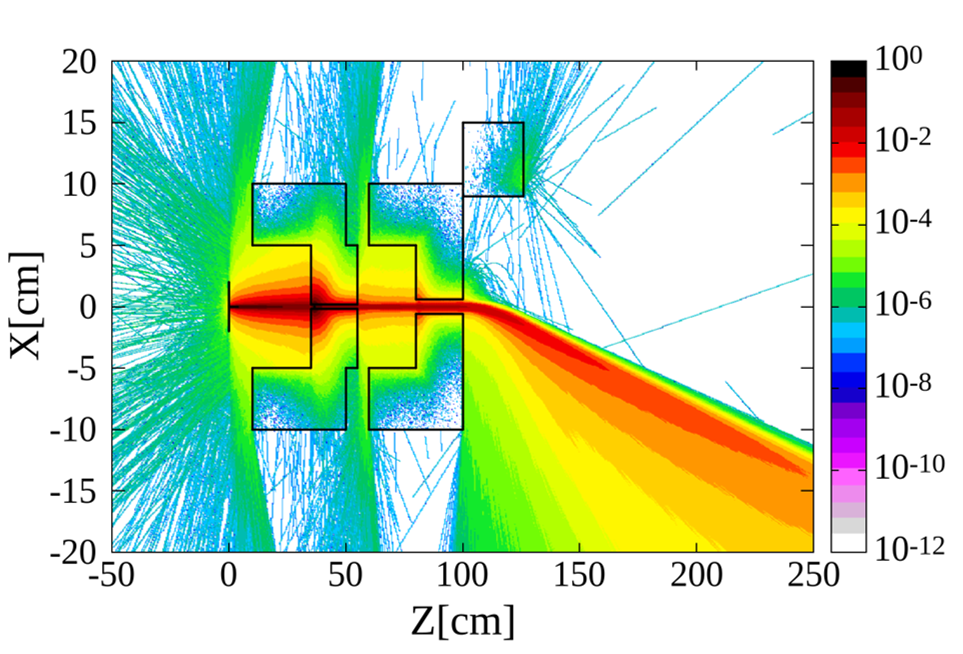}};
                \node[anchor=north west, font=\large\bfseries, yshift=0.3cm] at (image.north west) {(b)};
        \end{tikzpicture}
        \caption*{} 
        \refstepcounter{subfigure}\label{fig:posdis}
    \end{subfigure}
    \begin{subfigure}{0.5\textwidth} 
        \centering
        \begin{tikzpicture}
                \node[anchor=south west,inner sep=0] (image) at (0,0) {\includegraphics[width=\linewidth]{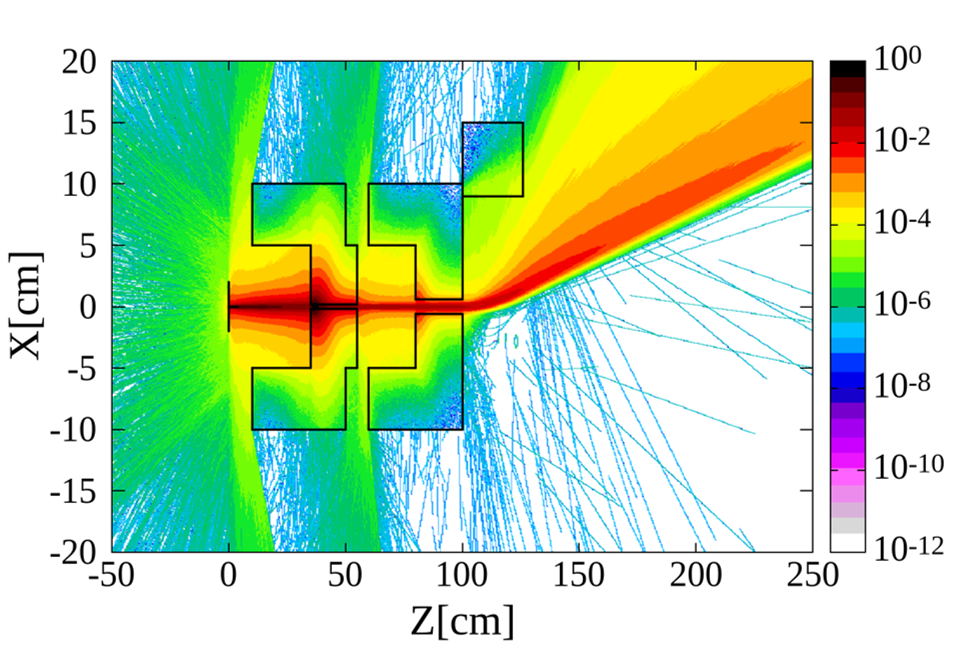}};
                \node[anchor=north west, font=\large\bfseries, yshift=0.3cm] at (image.north west) {(c)};
        \end{tikzpicture}
        \caption*{} 
        \refstepcounter{subfigure}\label{fig:eledis}
    \end{subfigure}\hfill 
    \begin{subfigure}{0.5\textwidth} 
        \centering
        \begin{tikzpicture}
                \node[anchor=south west,inner sep=0] (image) at (0,0) {\includegraphics[width=\linewidth]{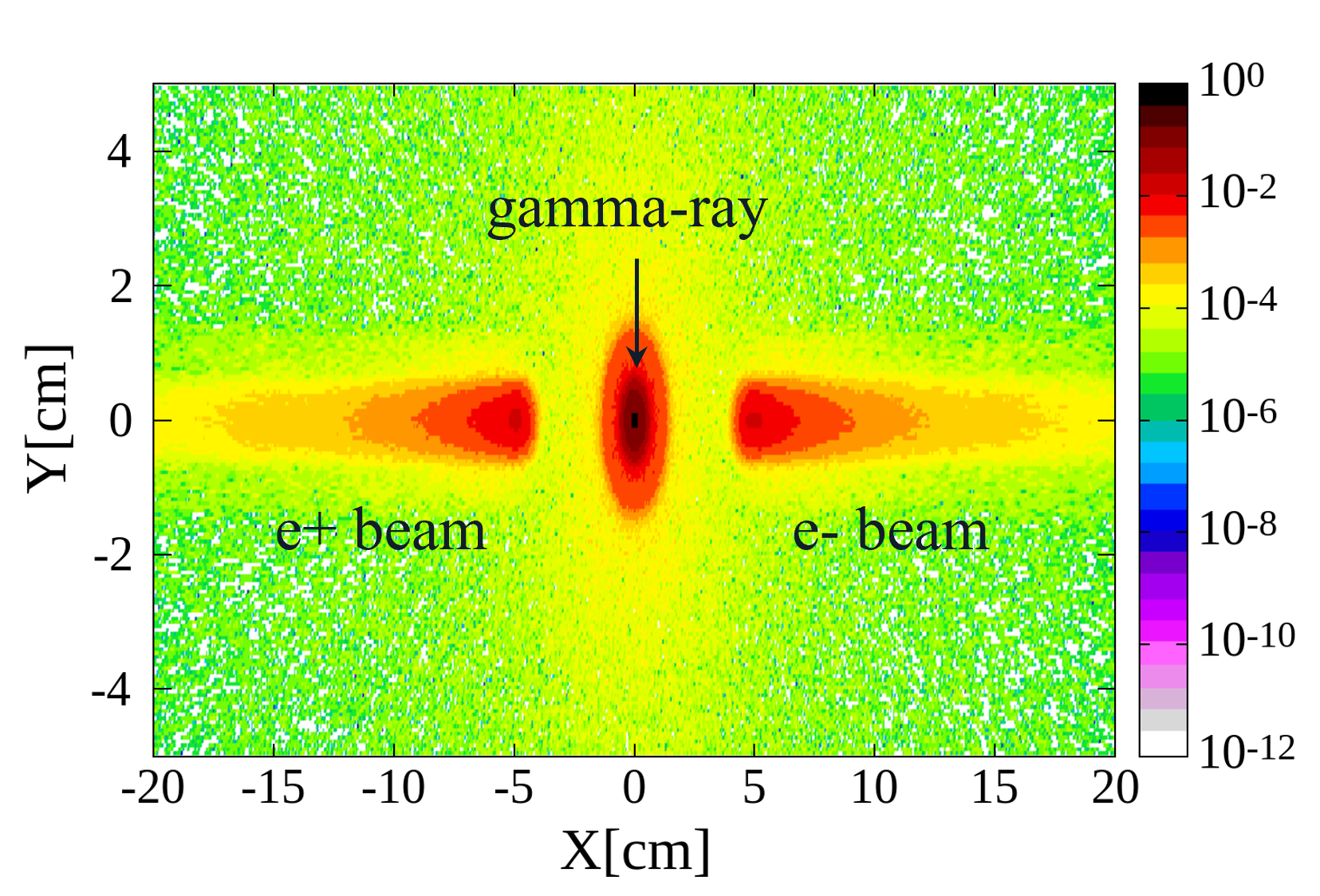}};
                \node[anchor=north west, font=\large\bfseries, yshift=0.3cm] at (image.north west) {(d)};
        \end{tikzpicture}
        \caption*{} 
        \refstepcounter{subfigure}\label{fig:ip_detect}
    \end{subfigure} 
    \vspace{-0.9cm}
    \caption{Time-integrated fluence distribution for (a) photons, (b) positrons, and (c) electrons resulting from the propagation of a monoenergetic gamma-ray beam. (d) The transverse distribution of electrons, positrons, and photons at the back of the spectrometer. The color bar represents particles/gamma photons per square centimeter.}
    \label{fig:distribution}
\end{figure*}
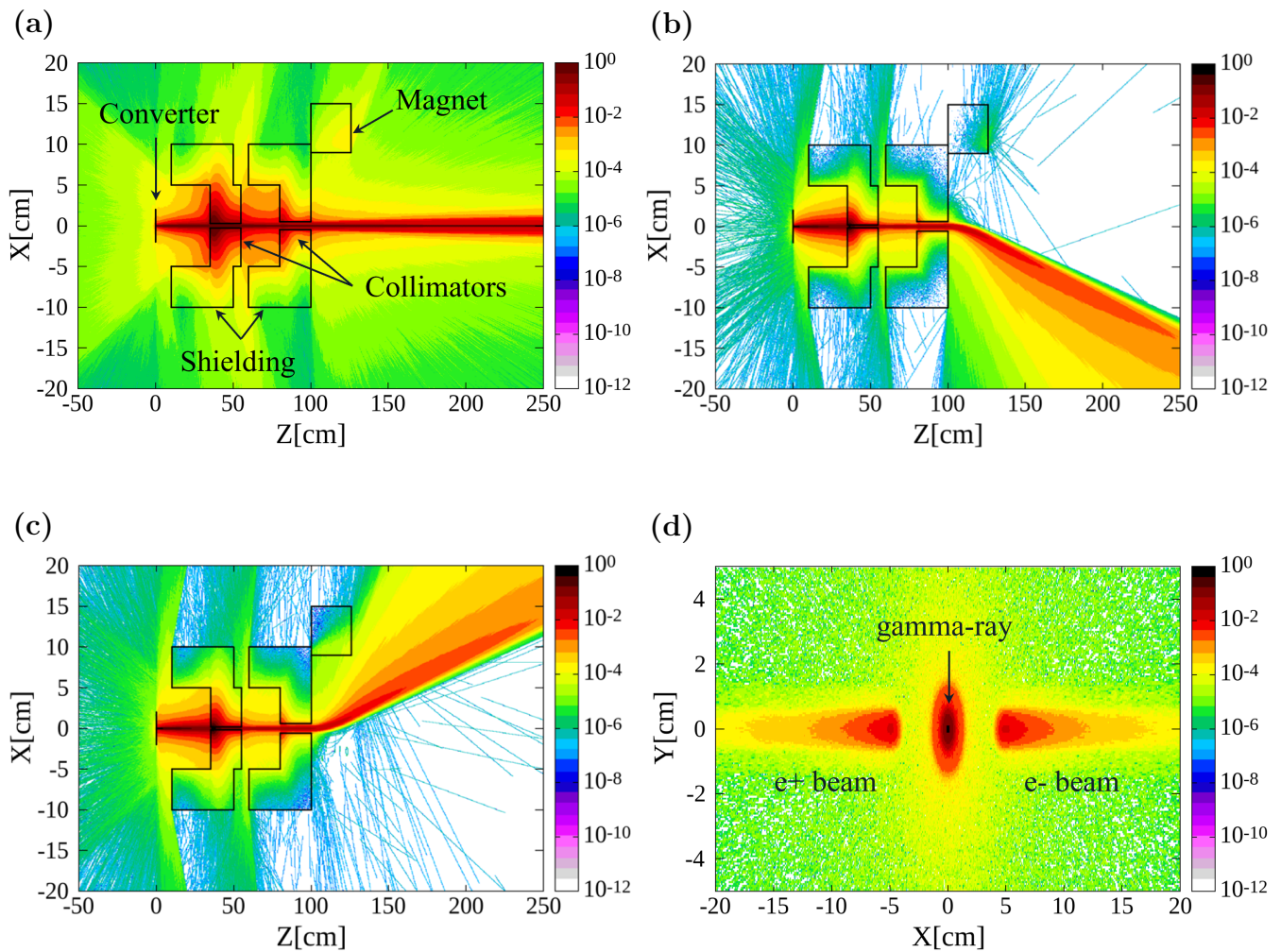

In order to study the overall performance of the spectrometer, Fig.\ref {fig:distribution} shows the two-dimensional particle distribution of $10^7$ photons propagating through the gamma-ray spectrometer simulated using the FLUKA code \cite{battistoniFLUKACodeDescription2007}. Lead shielding is added in front of the first collimator and between the double collimators to shield off-axis noise. Lead shielding must be placed axially symmetrically with the propagation axis of gamma-rays. The lead shielding in front of the first collimator in Fig.\ref{fig:1} is used to remove off-axis noise generated by the interaction of gamma-rays with the converter. Their apertures are 5.5 mm and 10 mm, forming an acceptance angle of 10 mrad with the source. The lead shielding between the double collimators is used to shield the noise generated by the first collimator. Since the radiation length of lead is approximately 0.56 cm \cite{nisttungsten_W}, increasing the transverse size of the shielding in our setup will not enhance the shielding effectiveness. The converter is set to a tungsten target, the collimator is set to lead material, and the entire system is set in a vacuum, neglecting the effects of pair production interacting with air \cite{abramowiczTechnicalDesignReport}. We set the thickness of the converter to 1 mm as previously optimized. The simulation includes the yoke of a C-shaped magnet that may be collided with by low-energy electrons deflected by the magnetic field. The magnetic field is set to 1 Tesla with dimensions of 2 × 8 × 32 cm (based on the actual magnet parameters). From Fig.\ref{fig:phodis}, \ref{fig:posdis}, \ref{fig:eledis} simulations demonstrate that after propagating through the spectrometer system, the distribution of electron-positron pairs at the detection plane exhibits a highly symmetric pattern, confirming effective collimation and a negligible off-axis noise contribution resulting from the optimized lead shielding geometry. In Fig.\ref{fig:eledis}, although collisions between low-energy particles and the magnet yoke produce noise spikes, their effect on detectors far from the yoke is negligible. When positrons traverse the detector plane, the electronic signals recorded thereon can be treated as ambient background noise. Fig.\ref{fig:ip_detect} shows the transverse distribution of electrons, positrons, and photons at the back of the spectrometer. The detectors are symmetrically positioned along the dispersion axis in the deflection paths of electrons and positrons. As shown in the figure, the detectors exhibit a signal-to-noise ratio $>$ 10, with signal intensities ranging from approximately $ 10^{-4}$ to $ 10^{-2}$ particles/gamma photon/cm$^{2}$. 

The energy resolution of a gamma-ray spectrometer is mainly limited by the resolution of the detector plane and the divergence of the incident particle beam \cite{cavanaghExperimentalCharacterizationSingleshot2023,fleck2025strong}. Taking factors such as the divergence angle $\Delta\theta_S$ of the positrons/electrons after conversion, detector spatial resolution $\delta x$(considering Fujifilm BAS-MS IP \cite{boutouxStudyImagingPlate2015,bonnetResponseFunctionsImaging2013}), and magnetic field inhomogeneity $B(z)$ into account, the energy resolution is 
\begin{equation}
  \frac{\delta  E}{E}=\frac{ E(L_S+L_M+L_D)\Delta\theta_S\oplus E\delta x  }{ce\int_0^{L_M}B(z)(L_M-z+L_D)dz}\label{resolution}
\end{equation}
Where $L_M$ is the length of the magnetic field, $L_S$ is the distance between the converter and the magnet entrance, $ c$ is the speed of light, and $L_D$ is the horizontal axial distance between the magnet exit and the detector. For our setup, $L_S=1$m, $L_M=32$cm, $L_D=70$cm, $\Delta\theta_S=10$mrad, $\delta x=25\mu$m, the magnetic field peak is 1.022 tesla. This could be expressed as $\frac{\delta  E}{E}$ = 0.24 * E[GeV]. For the 100 MeV particle, $\delta E/E$ is calculated as 2.4\%. Expanding the distance between the detector plane and the rear edge of the magnet, or increasing the longitudinal length of the shielding, could significantly reduce this value. Note that our design is effective for gamma-rays ranging from tens of MeV to a few GeV. For higher-energy gamma-rays, the longitudinal size of the collimator and lead shielding needs to be increased.
\section{Learning procedure of gamma spectrum deconvolution}\label{Learning procedure of gamma spectrum deconvolution}
For specially designed gamma-ray spectrometers, the inverse solution algorithm for obtaining gamma-ray spectra needs to be specially devised. To describe the methodology for reconstructing the gamma-ray beam spectrum, let $\mathcal{X}_{\gamma}\left(E_{\gamma}\right) $ be the energy distribution of the gamma-rays, and $ \mathcal{Y}_{l}\left(E_{l}\right)$ be the lepton pair (electron-positron pair) spectrum generated by gamma-rays interacting with converters. This transformation process can be defined as
\begin{equation}
\mathcal{Y}_{l}\left(E_{l}\right)=\int \mathcal{X}_{\gamma}\left(E_{\gamma}\right) \mathcal{A} _{\gamma l}\left(E_{\gamma}, E_{l}\right) d E_{\gamma}\label{eq:3}
\end{equation}
where $\mathcal{A} _{\gamma l}\left(E_{\gamma}, E_{l}\right) $ is the response matrix of the spectrometer system. Eq.\ref{eq:3} is essentially a Fredholm integral equation of the first kind \cite{reginattoOverviewSpectralUnfolding2010}, and solving the inverse problem belongs to the class of ill-posed problems. Using conventional methods to solve this inverse problem may result in a non-unique solution. Furthermore, slight noise introduced on the right side of the equation may cause significant errors in the solution \cite{burovaSolutionFredholmIntegral2021}. We need to perform precise inversion of the photon spectrum under high-noise conditions. Since Poisson noise and Gaussian noise affect the measurement process, the inverse problem  can be discretized into
\begin{equation}
y_i = A_{ij}x_j + \epsilon_i\label{eq:4}
\end{equation}
Here, the response matrix  $A_{ij}$ and the statistical noise $ \epsilon_i $ mainly from Poisson and Gaussian counts, are separated \cite{zhangDenoisingAutoencoderAided2021}. The advantage of this approach is to simulate the response matrix using the Geant4 MC     code by considering the geometric structure of the gamma-ray spectrometer and the experimental environment. The problem can then be decomposed into two steps: denoising and reconstruction.
\begin{figure}[htbp]
    \centering
    \includegraphics[width=1.0\linewidth]{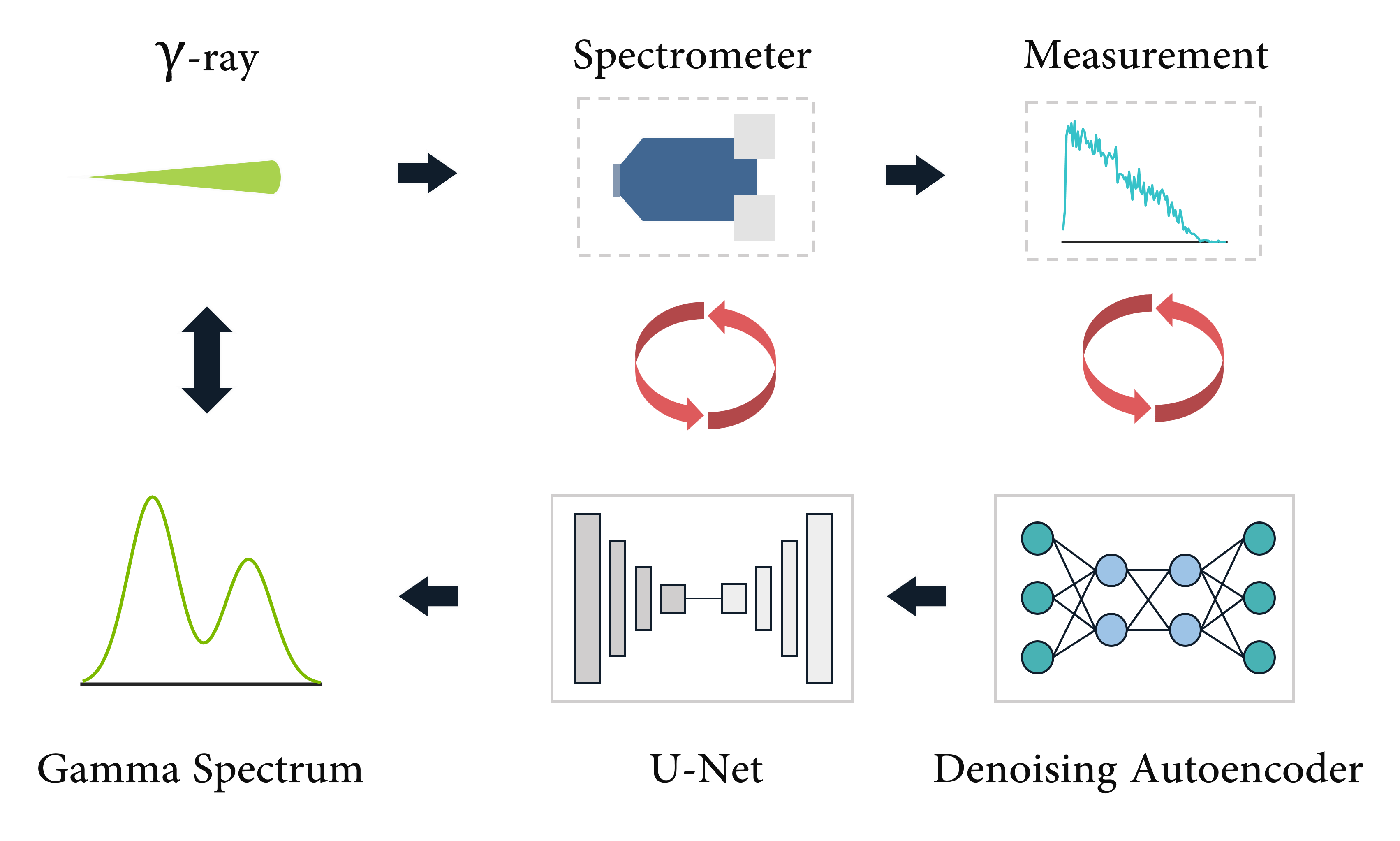}
    \caption{Schematic of the overall reconstruction workflow, organized into two sequential stages: (1) Denoising—a supervised autoencoder is trained on a hybrid dataset to suppress measurement noise while retaining key spectral features; (2) Deconvolution—the trained network, built on a U-Net architecture. This figure can be viewed as illustrating the forward and inverse processes of Eq.\ref{eq:4}.}
    \label{fig:whole}
\end{figure}

\begin{figure*}[htbp]
    \centering
    \begin{subfigure}{0.5\textwidth} 
        \centering
        \begin{tikzpicture}
                \node[anchor=south west,inner sep=0] (image) at (0,0) {\includegraphics[width=\linewidth]{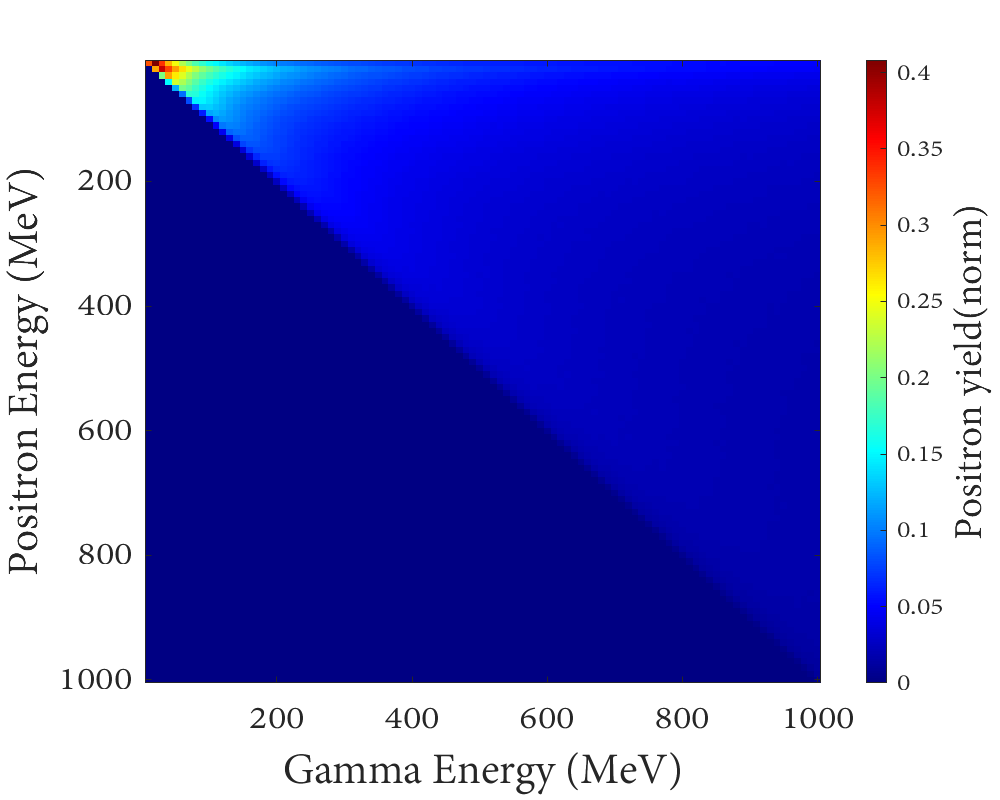}};
                \node[anchor=north west, font=\large\bfseries] at (image.north west) {(a)};
        \end{tikzpicture}
        \caption*{} 
        \refstepcounter{subfigure}\label{fig:Atarget_100um}
    \end{subfigure}
    \begin{subfigure}{0.5\textwidth}
        \centering
        \begin{tikzpicture}
                \node[anchor=south west,inner sep=0] (image) at (0,0) {\includegraphics[width=\linewidth]{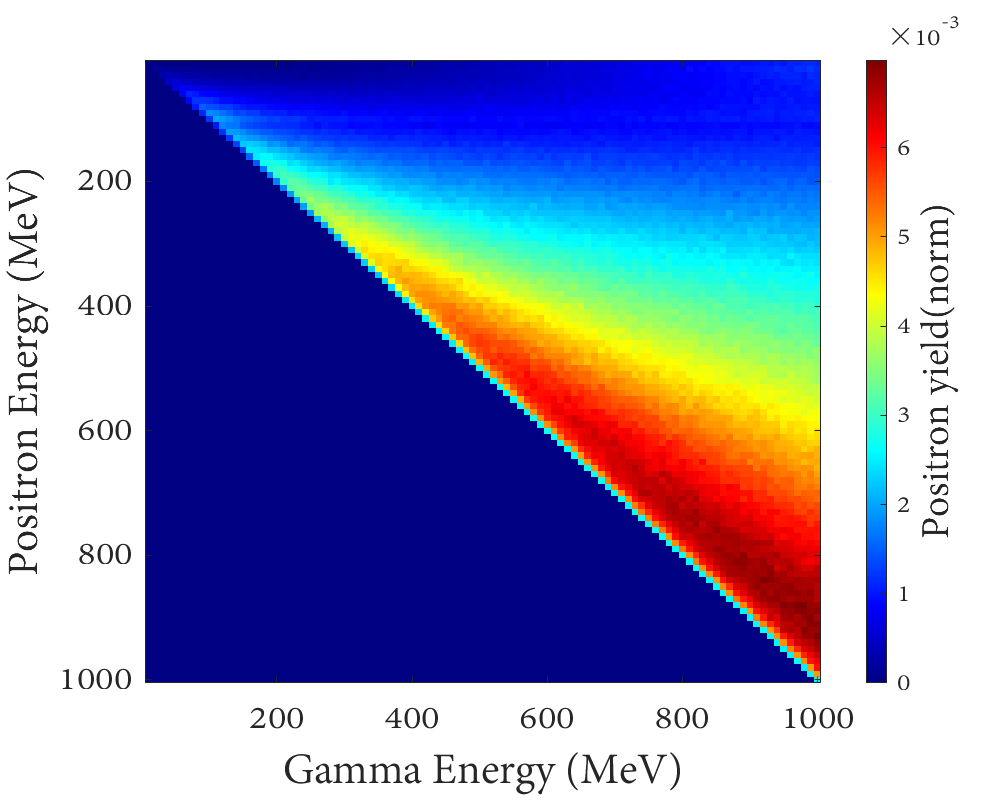}};
                \node[anchor=north west, font=\large\bfseries] at (image.north west) {(b)};
        \end{tikzpicture}
        \caption*{} 
        \refstepcounter{subfigure}\label{fig:Atarget_3mm}
    \end{subfigure}
    \vspace{-1cm}
    \caption{Multiple response functions were generated by simulating the monochromatic gamma-ray response using a gamma-ray spectrometer system configured as shown in Fig.\ref{fig:1} within the Geant4 framework. Vertical lineouts of this figure show the positron spectrum generated at a specific gamma-ray energy. (a) Ideal response of monochromatic gamma-rays interacting with a 1 mm tungsten converter. (b) Response function of a gamma-ray spectrometer with a 1 mm tungsten converter.}
    \label{fig:6}
\end{figure*}
The overall reconstruction workflow, depicted in Fig.\ref{fig:whole}, is organized into two sequential stages: denoising and deconvolution. In the denoising stage, we assemble a comprehensive training set composed chiefly of MC–simulated spectra supplemented by a subset of experimentally acquired data from the experimental data. This hybrid dataset is used to train a supervised denoising autoencoder, which learns to suppress measurement noise while preserving key spectral features. In the subsequent deconvolution stage, the trained network replaces conventional iterative solvers, such as the simultaneous iterative reconstruction technique (SIRT) \cite{hadenHighEnergyXray2020} or the Expectation Maximization Method (e.g., Richardson–Lucy algorithm \cite{liRichardsonLucyDeconvolutionMethod2019}) by performing paired-spectrum inversion in a single forward pass. Here, we employ a U-Net architecture, carefully tuning hyperparameters to optimize convergence and generalization. Below, we detail the procedure for generating synthetic spectra and characterizing the detector response matrix,  the design and configuration of the U-Net model, and the training regimen used to achieve robust, high-fidelity spectrum reconstruction.
\subsection{Dataset generation}\label{Dataset}
A comprehensive gamma-ray energy spectrum dataset is generated by combining over 30 different base spectra, typical radiation spectra, and experimental data from various sources (0.01 GeV–1 GeV). The base spectra include Gaussian distributions, exponential decay functions, step functions, and arbitrary random distributions across bins. Typical radiation spectra consist of monoenergetic spectra, bremsstrahlung spectra, synchrotron-like radiation spectra, and inverse Compton scattering spectra. Experimental data are derived from bremsstrahlung and inverse Compton scattering spectra generated in the experiments conducted at the Key Laboratory for Laser Plasmas (LLP) at Shanghai Jiao Tong University (SJTU) \cite{chen2024platform,liHighintensityLasersResearch2025}.

The spectrum parameters (e.g., peak position, width, intensity, Compton edge) are randomly sampled from a physically relevant range. To enhance the applicability of the data, data augmentation techniques are applied, including noise injection, random superposition, and spectrum-type manipulation. Gaussian-Poisson mixed noise with random amplitudes ranging from 0\% to 10\%(based on the worst-case signal-to-noise ratio of the spectrometer) is added to the spectra. Multiple spectra are linearly superimposed to form new spectra, generating realistic radiation patterns observed in experiments (e.g., double-peak or multi-peak structures) \cite{angioiNonlinearSingleCompton2016,heinzlLocallyMonochromaticApproximation2020}. Spectrum-type manipulation involves random scaling, shifting, and the injection of additional peaks.

Two priors are specifically considered: pair production must exceed 1.022 MeV, and the spectrum must remain non-negative. The augmented spectra are normalized in terms of photon number and subjected to non-negativity constraints. The final dataset consists of the original spectra $X$ and the noisy measurements $Y$ obtained after convolution with the detector response.
Details on constructing the response matrix $\mathcal{A}$ can be found in Sec. \ref{Building response matrix}.
\subsection{Building response matrix $\mathcal{A}$}\label{Building response matrix}
The gamma-ray spectrometer shown in Fig.\ref{fig:distribution} is reproduced in a Geant4 MC simulation. This model integrates precise detector geometry (including the magnet spectrometer section and the effect of air) and relevant physical processes (photoelectric effect, Compton scattering, and pair production) to characterize its response capabilities realistically. 

A series of monoenergetic gamma-ray sources, spanning the energy range of interest, is simulated to interact with the system. For each energy, we count the deposited energies across the corresponding detector channels, thereby populating columns of the discrete response matrix ${A}_{ij}$.
The simulated response matrix $\mathcal{A}$, shown in Fig.\ref{fig:6}, consists of two panels that illustrate the interaction of monochromatic gamma-rays with different systems. The heatmaps present the received energy on the detector plane versus the incident gamma-ray energy, highlighting how the system responds to various gamma-ray energies. In the discrete representation $A_{ij}$, each column $j$ encodes the probability distribution of positron energies $\{E_l^i\}$ arising from a monoenergetic gamma-ray with energy $E_\gamma^j$. This results in a specialized Fredholm-type equation to a Volterra integral equation of the first kind, which has a variable limit of integration. This has advantages and disadvantages for the framing of the inverse problem. The upper-triangular nature of the heatmaps is indicative of the causal constraint, where non-zero entries appear only when the incident photon energy exceeds the positron energy.
\begin{figure*}[hbtp]
    \centering   
    \begin{subfigure}{0.427\textwidth}
        \centering
        \begin{tikzpicture}
                \node[anchor=south west,inner sep=0] (image) at (0,0) {\includegraphics[width=\linewidth]{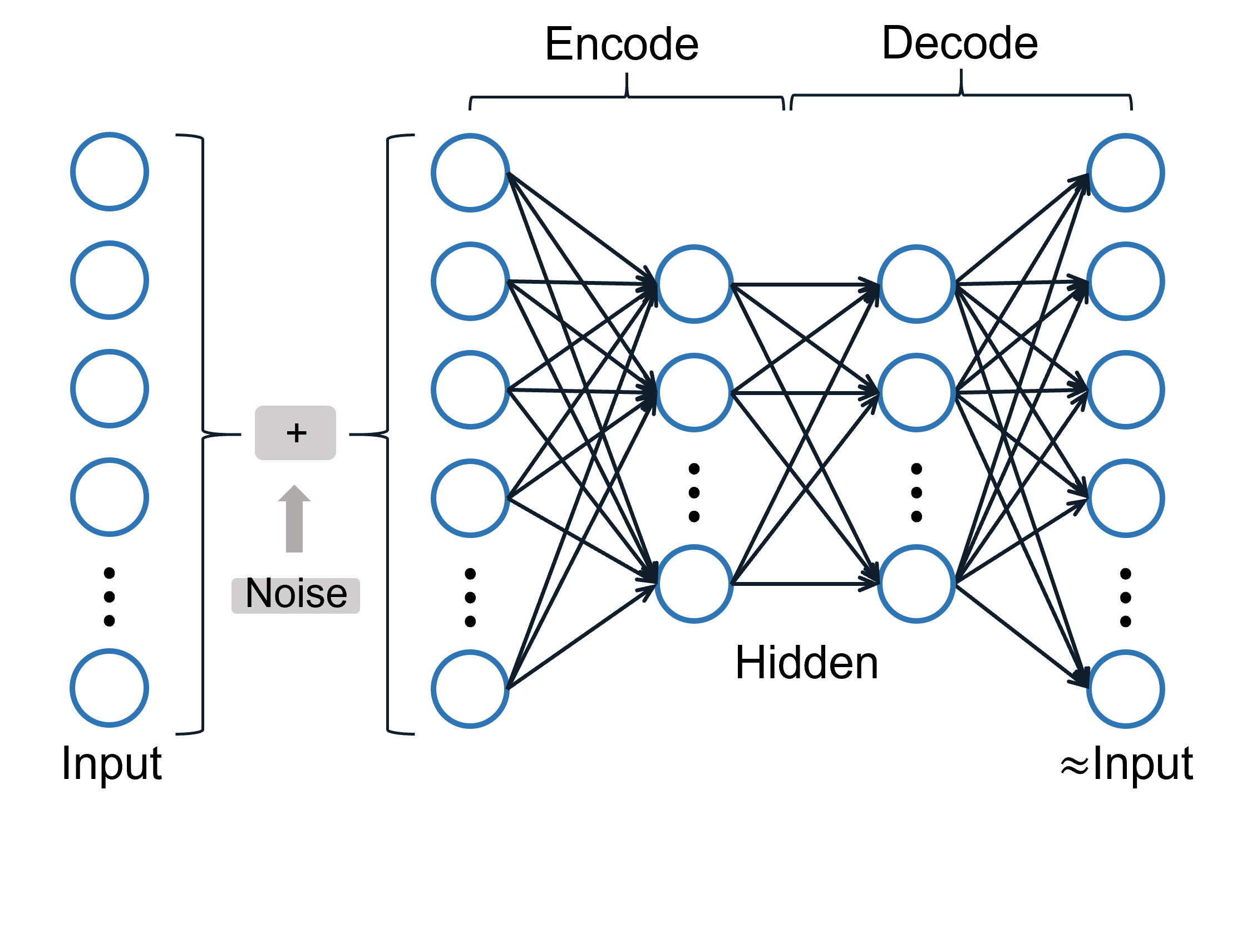}};
                \node[anchor=north west, font=\large\bfseries ,yshift = 7.95pt] at (image.north west) {(a)};
        \end{tikzpicture}
        \caption*{} 
        \refstepcounter{subfigure}\label{fig:dae}
    \end{subfigure}\hfill 
    \begin{subfigure}{0.56\textwidth}
        \centering 
        \begin{tikzpicture}
                \node[anchor=south west,inner sep=0] (image) at (0,0) {\includegraphics[width=\linewidth]{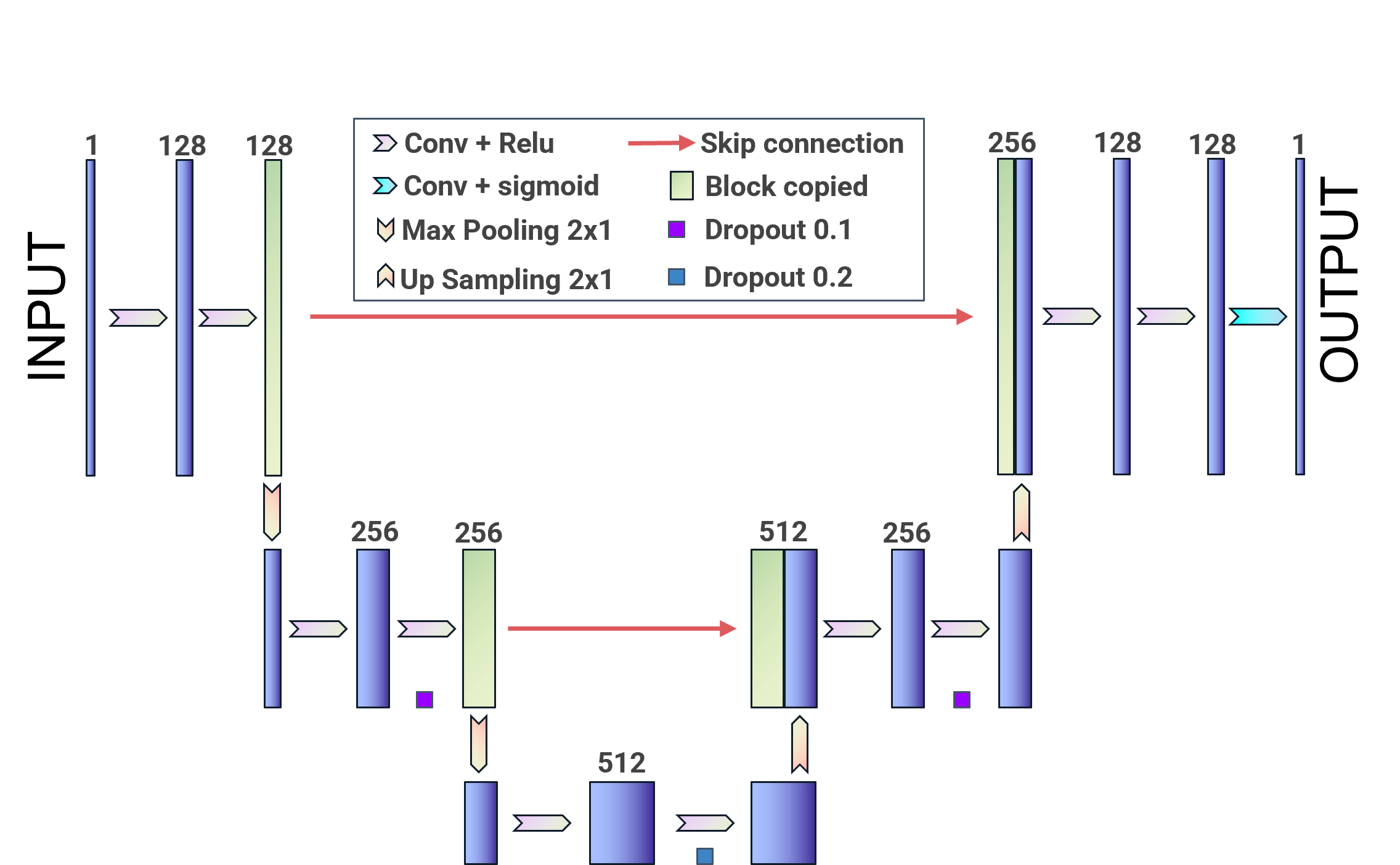}};
                \node[anchor=north west, font=\large\bfseries] at (image.north west) {(b)};
        \end{tikzpicture}
        \caption*{} 
        \refstepcounter{subfigure}\label{fig:unet}
    \end{subfigure}
    \vspace{-0.6cm}
    \caption{Schematic of machine learning architecture: (a) Denoising autoencoder (DAE) with symmetric encoder-decoder structure for suppressing statistical noise in measured positron spectra. The encoder compresses the input through dense layers to a latent representation, and the decoder reconstructs a noise-suppressed spectrum (b) U-Net deconvolution network for solving the inverse problem to recover incident gamma-ray spectra from denoised measurements.}
    \label{fig:ML}
\end{figure*}
This triangular architecture results in a sharp decay in the singular-value spectrum of $\mathcal{A}$, producing an exceedingly large condition number. Inversion by the least squares method therefore amplifies any measurement noise—whether Poisson or Gaussian—in the small singular value modes, producing non-physical oscillations or negative features in the recovered spectrum.

The ill-conditioning of the response matrix, characterized by a rapidly decaying singular-value spectrum and minimal channel overlap \cite{hansenDiscreteInverseProblems2010}, means that any conventional inverse solver—even with standard Tikhonov or truncated Singular Value Decomposition (SVD) regularization—inevitably sacrifices either resolution or stability. Moreover, the photon counting statistics introduce heteroscedastic noise that cannot be uniformly damped by a single global regularization parameter. These limitations motivate the adoption of a data-driven inversion strategy.
\subsection{Denoising autoencoder  for statistical‐noise suppression}\label{Denoising autoencoder}
To decouple intrinsic spectral features from photon-counting fluctuations, we introduce a supervised denoising autoencoder (DAE) that learns a direct mapping from noisy measurements $Y$ to noise-attenuated spectra $\widetilde{Y}$ to make the features robust. DAEs have already been used as a preprocessing technique for specific spectrometer systems, similar to Zhang et al.'s preprocessing technique \cite{zhangDenoisingAutoencoderAided2021}, where the original measurement data is denoised before reconstruction, and then the spectrum is reconstructed based on the denoised data. Fig.\ref{fig:dae} presents a schematic of the DAE architecture, highlighting its encoder–decoder symmetry structure optimized for one-dimensional spectral data:
\begin{itemize}
  \item \textbf{Encoder:} The input spectrum of length \(N\) is first flattened and then passed through two successive dense–ReLU layers that compress the representation from \(N\) to a first hidden-layer dimension \(h_1\) and finally to the latent hidden-layer dimension \(h_2\), thereby isolating salient spectral features while attenuating stochastic noise.
  \item \textbf{Decoder:} The latent vector of dimension \(h_2\) is symmetrically expanded via a ReLU‐activated dense layer back to \(h_1\), followed by a linear dense layer that reconstructs the full-length \(N\) output, which is then reshaped to the original \((N,1)\) spectrum.
\end{itemize}
Training is performed on 500000 paired spectra (see, in Section \ref{Dataset}) containing $X_{\mathrm{clean}},Y_{\mathrm{noisy}}$, where $X_{\mathrm{clean}}$ spans analytic base shapes, MC–simulated mixtures, and a subset of experimental measurements, and $Y_{\mathrm{noisy}}=A\,X_{\mathrm{clean}}+\epsilon$ incorporates both Poisson and Gaussian perturbations at amplitudes up to 10\%. We optimize the mean-squared error between the autoencoder output and $X_{\mathrm{clean}}$ using the Adam optimizer (learning rate $10^{-3}$). During each epoch, fresh noise realizations and random amplitude/shifting augmentations ensure robust generalization across count rates and spectral shapes.

\begin{figure*}[htbp]
    \centering
    \includegraphics[width=0.9\linewidth]{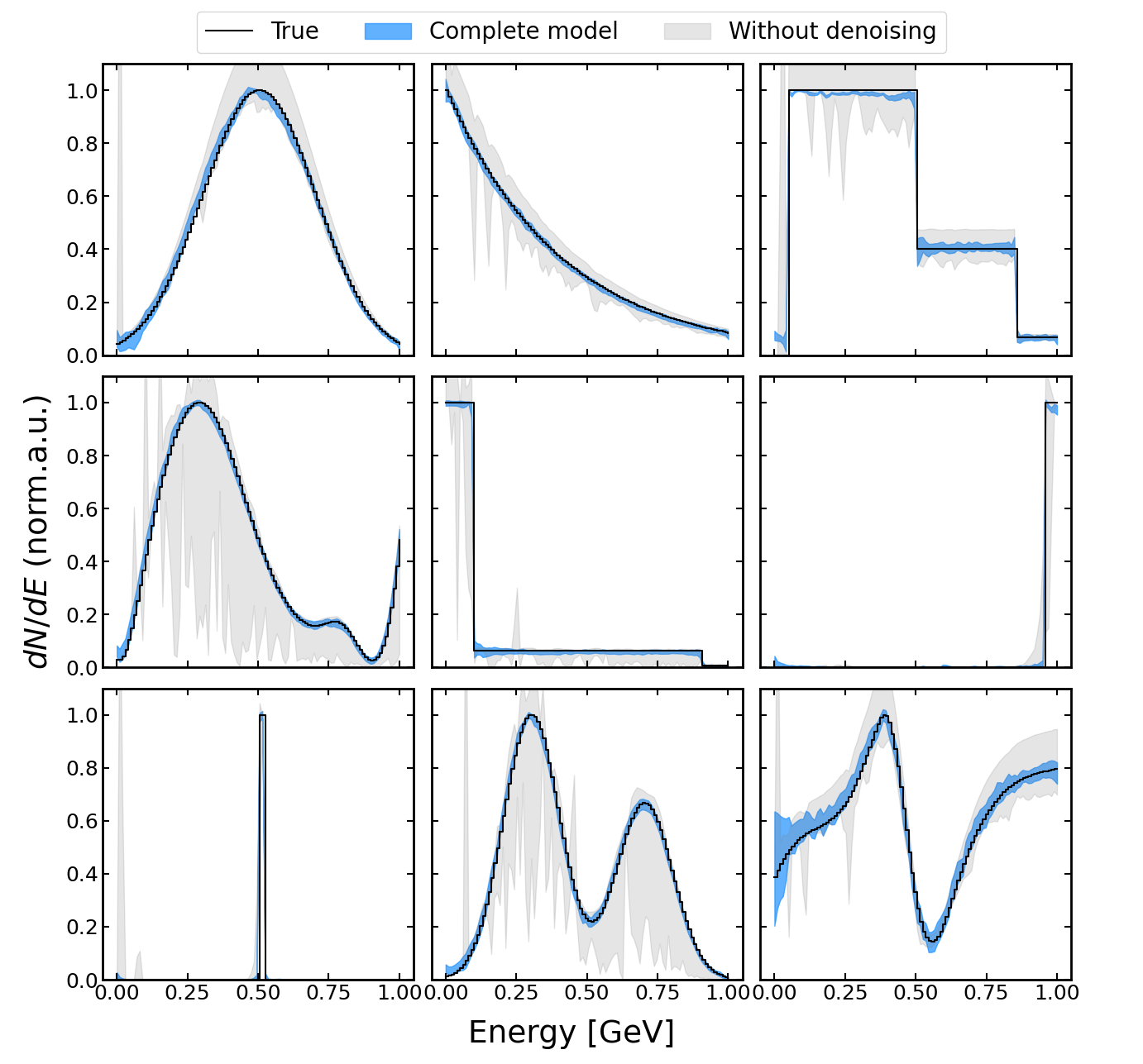}
    \caption{The reconstructed photon spectrum obtained by applying machine learning algorithms to the dataset(For the convenience of presentation, all the data has been normalized to the maximum value). The black line shows the photon spectrum incident on the spectrometer in the simulation. The blue shaded band indicates the 95\% Bayesian credible interval (Monte Carlo uncertainty) calculated by the algorithm. The shaded gray represents the result of deconvolution without denoising.}
    \label{fig:reconstruction result}
\end{figure*}

On a held‐out test set of 50000 spectra, the DAE achieves an average signal-to-noise ratio (SNR) of 30.2 dB and a mean absolute error below 10\% compared to the directly deconvolved data without denoising. As can be seen later, the role of DAE is indispensable in high-noise conditions (see Fig.\ref{fig:reconstruction result}). By suppressing stochastic noise to levels suitable for downstream inversion and quantitative analysis, we substantially improve the stability and resolution of the subsequent U‐Net–based inversion. 
\subsection{U-Net deconvolution}\label{U-Net deconvolution}
To address the ill-posed nature of the inverse problem in gamma-ray spectrum reconstruction, we employ a U-Net architecture, which is well-suited for deconvolution tasks, particularly in the presence of noise. The U-Net is a convolutional neural network designed for pixel-level prediction \cite{huDeepLearningAssisted2020}, and it is ideal for reconstructing the high-dimensional gamma-ray spectrum from noisy lepton pair measurements \cite{döppDatadrivenScienceMachine2023}.
The U-Net architecture consists of an encoder-decoder structure with skip connections, which facilitates the preservation of fine spectral features during the reconstruction process \cite{ronnebergerUNetConvolutionalNetworks2015}. The encoder compresses the input noisy spectrum by applying a series of convolutional layers with downsampling, capturing global spectral characteristics. The decoder then expands the compressed representation, recovering the original spectrum resolution while preserving local details through skip connections. These connections transfer information from the encoder to the decoder, ensuring high-fidelity reconstruction.
\begin{figure*}[htbp]
    \centering
    \begin{subfigure}{0.333\textwidth} 
        \centering
        \begin{tikzpicture}
                \node[anchor=south west,inner sep=0] (image) at (0,0) {\includegraphics[width=\linewidth]{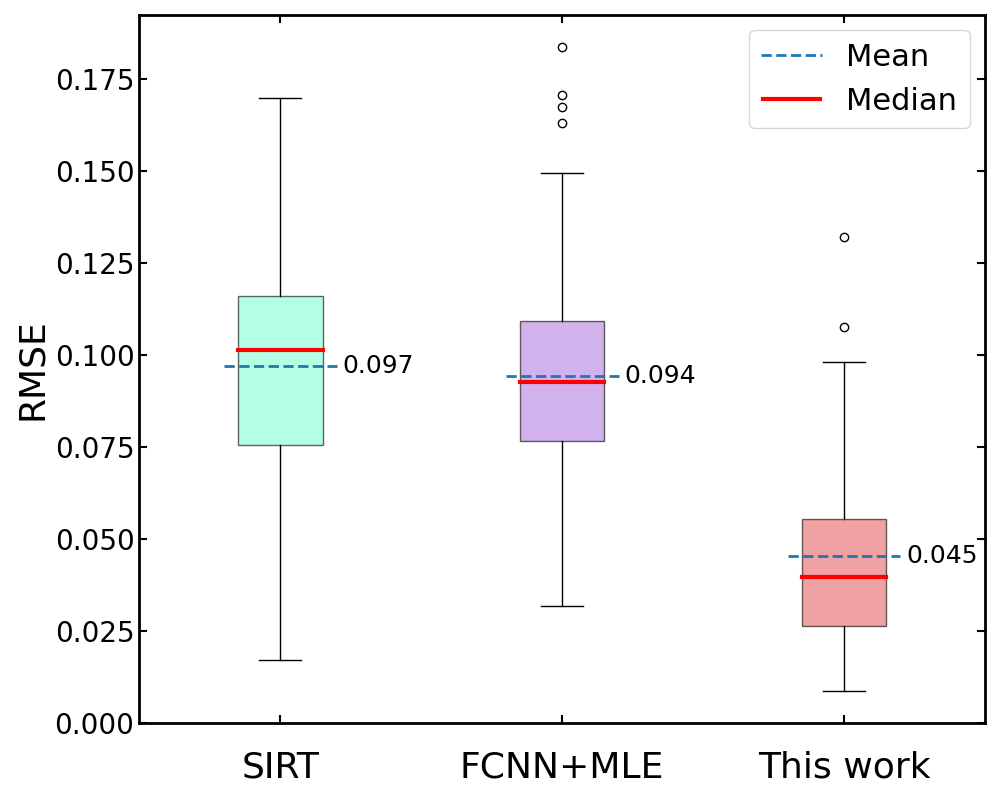}};
                \node[anchor=north west, font=\large\bfseries , yshift=15pt] at (image.north west) {(a)};
        \end{tikzpicture}
        \caption*{} 
        \refstepcounter{subfigure}\label{fig:RMSE}
    \end{subfigure}\hfill 
    \begin{subfigure}{0.333\textwidth}
        \centering
        \begin{tikzpicture}
                \node[anchor=south west,inner sep=0] (image) at (0,0) {\includegraphics[width=\linewidth]{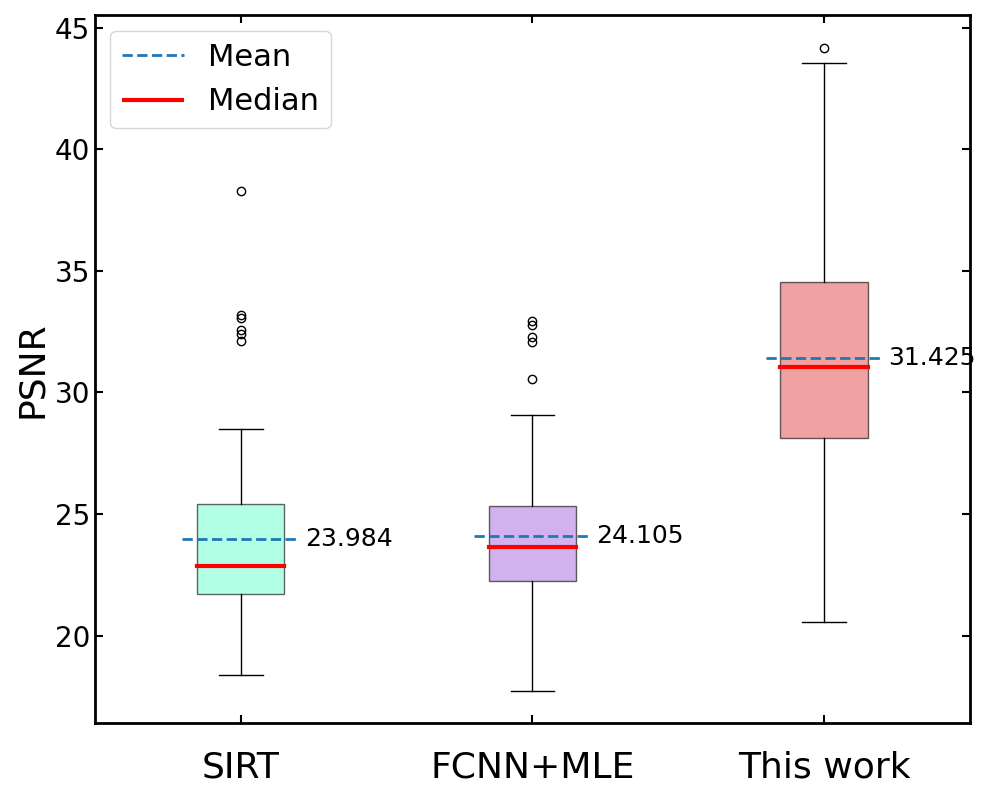}};
                \node[anchor=north west, font=\large\bfseries, yshift=15pt] at (image.north west) {(b)};
        \end{tikzpicture}
        \caption*{} 
        \refstepcounter{subfigure}\label{fig:PSNR}
    \end{subfigure}\hfill 
    \begin{subfigure}{0.333\textwidth} 
        \centering
        \begin{tikzpicture}
                \node[anchor=south west,inner sep=0] (image) at (0,0) {\includegraphics[width=\linewidth]{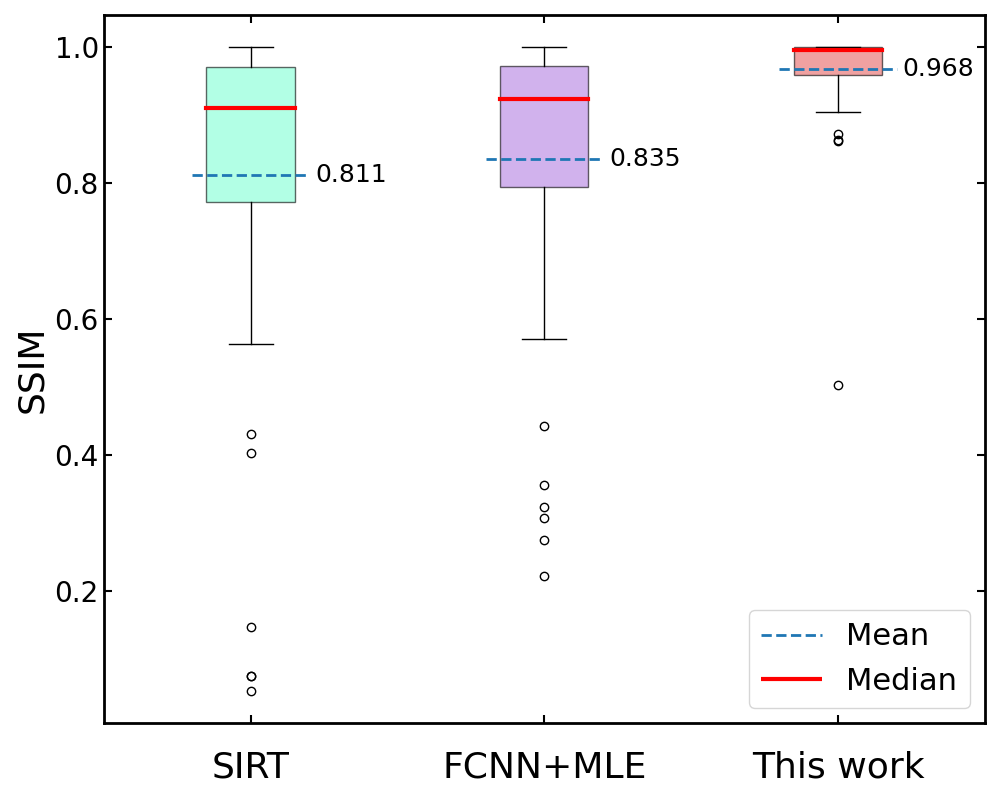}};
                \node[anchor=north west, font=\large\bfseries, yshift=15pt] at (image.north west) {(c)};
        \end{tikzpicture}
        \caption*{} 
        \refstepcounter{subfigure}\label{fig:SSIM}
    \end{subfigure}
    \vspace{-1cm}
    \caption{Performance of different models: (a) RMSE; (b) PSNR; (c) SSIM for the SIRT algorithm, FCNN+MLE model, and our deep learning approach. The boxplots demonstrate the distribution of each metric over 300 test spectra. For RMSE, lower values indicate better performance, while for PSNR and SSIM, higher values are desirable. The red line represents the median value. The blue dashed line represents the average value of the data. }
    \label{fig:evaluation}
\end{figure*}
This network is summarized in Fig.\ref{fig:unet}.  The encoder starts with a 1D convolutional layer, followed by ReLU activations and max-pooling operations. These layers progressively reduce the dimensionality of the input while learning to capture the spectral shape and intensity distribution, which are crucial for distinguishing between gamma-ray sources.
The bottleneck consists of the most abstract representation of the spectrum, where the spatial resolution is minimized, and the model learns the most essential features for reconstruction. This compressed form encodes the core information necessary for accurate spectrum recovery.
The decoder progressively upscales the feature map through transposed convolutions, effectively reconstructing the input spectrum. Skip connections are utilized to bring high-resolution features from the encoder directly to the corresponding layers in the decoder, ensuring that critical spectral details are retained.
 The final output layer produces the reconstructed gamma-ray spectrum, ensuring that it satisfies the physical constraints, such as non-negativity.

The U-Net was implemented using a Keras framework with a TensorFlow backend \cite{abadi2016tensorflow}, and the model is trained to minimize the Mean Squared Error (MSE) between the reconstructed spectrum and the true clean spectrum. The loss function used is:
\begin{equation}
    \mathcal{L}(\theta) = \frac{1}{N} \sum_{i=1}^{N} \left( \mathcal{Y}_i - \hat{\mathcal{Y}}_i \right)^2
\end{equation}
where $\mathcal{Y}_i$ represents the true clean spectrum, $\hat{\mathcal{Y}}_i$ is the predicted spectrum, and $N$ is the total number of samples in the batch. The model is trained using the Adam optimizer with a learning rate of $10^{-3}$ \cite{kingma2014adam}, which ensures efficient convergence while preventing overfitting to noise. The learning rate is a crucial hyperparameter, as it determines the speed at which the model updates the weights during training. An improperly chosen learning rate can lead to either slow convergence or overshooting the optimal solution.

In addition to the learning rate, key hyperparameters include the number of convolutional layers, the kernel size (set to 9 or 7), and the number of filters (128, 256, 512 in successive layers). Large kernel sizes can effectively capture the overall shape of spectral features, helping the network avoid interference from local statistical fluctuations. The first layer employs 128 channels with a convolution width of approximately 100 units, using a kernel size of 9. With each subsequent downsampling step, the number of channels is doubled, while the width is halved, accompanied by a reduction in kernel size. The narrowest convolution width reaches around 25 units, with the deepest layers achieving up to 512 channels. During training, the batch size is typically set to 64 or 128 to facilitate efficient training across varying sample sizes. These settings enable the model to capture both fine and coarse spectral features. Dropout regularization is applied at various stages to stabilize training and enhance the model’s generalization across different spectra.

Training is performed on a dataset mentioned in \ref{Dataset}. The final dataset consists of 500000 pairs of clean and noisy spectra, used for training the network. To restore the original scale of the reconstructed gamma spectrum, it is crucial to address the relationship between the photon spectrum and its integral normalization. The photon spectrum is essentially the product of the photon count and the normalized differential energy spectrum $\frac{dN}{dE}$. In this framework, the U-Net model learns the mapping between the normalized outputs $Y$ and $X$. The activation function of the output layer is sigmoid, which maps $X$ into the range $[0, 1]$. To recover the original energy spectrum, the output of the model must be subjected to a final integral normalization. Experimentally, the original gamma photon spectrum can be derived by using the photon counts from imaging plates and the normalized energy spectrum obtained through simulations.
\section{evaluation results}\label{evaluation results}
The results of the deconvolution process are shown in Fig.\ref{fig:reconstruction result}, where nine representative spectra with different parameters from the database were used to test the trained model. The black line is the original gamma-ray spectrum, the shaded gray represents the result of deconvolution without denoising, and the light blue confidence band represents model uncertainty. This figure also illustrates that the denoising process and the deconvolution process are closely related. Under high-noise conditions, the deconvolution process fails to accurately capture certain features of the original spectrum, and the impact of Poisson noise is significant. The uncertainty is employed as a measure of the model's confidence in its predicted values, quantified through MC dropout sampling. This process can be understood through the equivalence between dropout training and variational inference \cite{galDropoutBayesianApproximation2016}, where dropout acts as a regularization technique and is equivalent to Bayesian inference of the weight distribution.
During spectrum reconstruction ${y}^* \rightarrow {x}^*$, we sample $T$ weight configurations via dropout \cite{galDropoutBayesianApproximation2016}:
\begin{equation}
\mathbb{E}[{x}^*] \approx \frac{1}{T}\sum_{t=1}^T f^{{W}_t}({y}^*), \quad {W}_t \sim q\theta({W})
\end{equation}
Where ${y}^*$ is the input lepton spectrum, ${x}^*$ is the reconstructed spectrum, $\mathbb{E}[{x}^*]$ represents the expected reconstruction computed by averaging over multiple weight configurations ${W}_t$ sampled from the Bayesian distribution $q\theta({W})$.
It is viable to calculate the model’s predictive variance:
\begin{equation}
\sigma = \sqrt{\frac{1}{T}\sum_{t=1}^T \left( f^{{W}_t}({y}) - \mathbb{E}[{x}] \right)^2}
\end{equation}
Although not shown, the reconstruction of the electron spectrum produces similar results and can be used as a consistency check \cite{cavanaghExperimentalCharacterizationSingleshot2023}. The spectral shape of the incident photon spectrum is well reconstructed within the 95\% Bayesian credible interval of the reconstruction results.

The 95\% Bayesian Credible Interval (BCI) presented in this work is quantitatively defined as follows. For each energy channel \(i\) of the reconstructed spectrum, we compute the posterior distribution of the predicted flux value, \(p(x^*_i | {y}^*)\), based on \(T\) stochastic forward passes using MC dropout. The 95\% BCI for channel \(i\) is then defined as the equal-tailed interval bounded by the \(2.5{\text{th}}\) and \(97.5{\text{th}}\) percentiles of this empirical posterior distribution. Mathematically, the interval \( [L_i, U_i] \) satisfies:
\begin{equation}
 P(L_i \leq x^*_i \leq U_i | {y}^*) = 0.95   
\end{equation}
where \(L_i\) and \(U_i\) are the lower and upper bounds of the credible interval for the \(i\)-th energy channel, respectively. That means, given the observed lepton spectrum \({y}^*\), there is a 95\% probability that the true value of the incident photon flux in channel \(i\) lies within the interval \([L_i, U_i]\). The light blue confidence band in Fig.\ref{fig:reconstruction result} is generated by connecting the \(L_i\) and \(U_i\) values across all energy channels, providing a visual representation of the model's uncertainty over the entire reconstructed spectrum.  

The performance of gamma-ray energy spectrum reconstruction is evaluated using three metrics: root mean square error (RMSE), peak signal-to-noise ratio (PSNR), and structural similarity (SSIM) \cite{mostafapourFeasibilityDeepLearningGuided2021}.
RMSE is a classic statistical measure of the average deviation between the reconstructed spectrum and the reference spectrum. It is defined as the square root of the mean of the sum of the squares of the errors in all energy channels:
\begin{equation}
\mathrm{RMSE} = \sqrt{\frac{1}{N}\sum_{i=1}^N \bigl(\hat{x}_i - x_i\bigr)^2}
\end{equation}
where $x_i$ and $\hat{x}_i$ represent the true count and reconstructed count for the $i$th channel, respectively, and $N$ is the total number of channels. RMSE is sensitive to large deviations and increases significantly when the reconstruction overestimates or underestimates the counts in certain energy channels, thereby reflecting the overall reconstruction accuracy at a rough level.
PSNR originates from the field of image processing and is used to measure the contrast strength between signals and noise on two spectral lines. Its formula is typically written as
\begin{equation}
\mathrm{PSNR} = 10 \log_{10}\!\Bigl(\frac{\mathrm{MAX}^2}{\mathrm{MSE}}\Bigr)
\end{equation}
where $\mathrm{MAX}$ is the maximum count value of the true spectral line, and $\mathrm{MSE}$ is the square of the aforementioned RMSE. PSNR is measured in decibels (dB), with higher values indicating that the reconstructed spectral line retains more of the original signal features overall, and the relative contribution of noise and reconstruction error is reduced.
SSIM focuses on preserving the local structure and contrast details of the spectral line. Its core idea is that a high-similarity reconstruction should not only be close in amplitude but also consistent in shape, contrast, and brightness distribution. For spectral line segment $x$ and reconstructed segment $y$, SSIM is typically defined as
\begin{equation}
\mathrm{SSIM}(x,y) = \frac{(2\mu_x \mu_y + C_1)(2\sigma_{xy} + C_2)}{(\mu_x^2 + \mu_y^2 + C_1)(\sigma_x^2 + \sigma_y^2 + C_2)}
\end{equation}
where $\mu$ and $\sigma$ denote the local mean and variance, respectively, $\sigma_{xy}$ is the covariance. The constants $ c_{1}=\left(k_{1} L\right)^{2}$,$ c_{2}=   \left(k_{2} L\right)^ {2}$   are small constants used to stabilise the denominator, where L is the dynamic range of the spectral values (i.e., the maximum value minus the minimum value), typically taken as  $k_{1}=0.01$, $k_{2}=0.03$.  The SSIM value ranges from $[0,1]$. The closer it is to 1, the more consistent the reconstructed spectral line is with the reference spectral line in terms of detail and overall structure.

To demonstrate performance, we reproduced the configurations of SIRT and the hybrid approach using fully connected neural networks to generate an informed initial guess for iterative maximum likelihood estimation (MLE) \cite{hadenHighEnergyXray2020,yadavReconstructingGammarayEnergy2024}. The statistical results of 300 test datasets (average signal-to-noise ratio approximately 35.70 dB) evaluated using three methods are shown in Fig.\ref{fig:evaluation}. Clearly, our method achieved the lowest RMS error and the highest peak signal-to-noise ratio and structural similarity in the comparison.
\section{Conclusion and Outlook}\label{Conclusion and Outlook}
In this study, we have developed a transformative approach for high-fidelity gamma-ray spectroscopy in the multi-MeV to GeV range, addressing critical challenges in spectral reconstruction through the integration of optimized spectrometer parameters and an advanced machine learning deconvolution algorithm. It should be stressed that this type of spectrometer is almost transparent to the gamma rays, with only a small percentage of them interacting. It can thus be used in conjunction with other gamma-ray detectors (profilers and polarimeters). We propose a machine learning framework, which combines a denoising autoencoder for statistical-noise suppression with a U-Net architecture for spectral deconvolution. Denoising autoencoders, trained specifically to remove 90\% or more of the input noise, function analogously to a smoothing operation. Also, U-Net architectures demonstrate remarkable robustness, effectively reconstructing the original energy spectrum even from partially noisy data. This approach effectively solves the ill-posed inverse problem inherent in gamma-ray spectroscopy, overcoming limitations of traditional algorithms such as SIRT and hybrid methods.  

Validation against a diverse dataset of synthetic and experimental spectra demonstrated superior performance, with our method achieving the lowest root mean square error (RMSE) and highest peak signal-to-noise ratio (PSNR) and structural similarity (SSIM) among compared techniques. The ML architecture's ability to preserve spectral features while suppressing noise enables robust reconstruction of complex spectral shapes, including those with multi-peak structures characteristic of nonlinear Compton scattering and LWFA-driven bremsstrahlung.  

The present spectrometer is designed to operate robustly for photon energies $\gtrsim$50 MeV up to the order of a few GeV. In future strong-field laser physics experiments, extending the energy range to several GeV or more would require enlarging spectrometer components and optimizing noise shielding to account for higher-order quantum electrodynamics (QED) processes. Further work will cover multi-parameter reconstruction (e.g., the angular distribution of incident photons). The ability to extract the ``double-differential'' of the photon beam (i.e., $\frac{d^2N}{dEd\theta}$) would be great for SFQED experiments. This work establishes a foundation for next-generation gamma-ray diagnostics, with implications for fundamental SFQED studies, laboratory astrophysics, and compact radiation sources.  
\section*{Acknowledgement}
This work was supported by the National Key Research and Development Program of China(Grant No. 2021YFA1601700), Al for Science Program, Shanghai Municipal Commission of Economy and Informatization (Grand No. 2025-GZL-RGZN-BTBX-02029), the National Natural Science Foundation of China (Grant No. 12074251), the Strategic Priority Research Program of the Chinese Academy of Sciences(Grants No. XDA25010500, No. XDA25010100), and the Engineering and Physical Sciences Research Council (EPSRC) (Grant EP/V049186/1). The authors would like to acknowledge the sponsorship from the Yangyang Development Fund. The computations in this paper were run on the Siyuan-1 cluster supported by the Center for High Performance Computing at Shanghai Jiao Tong University.

\bibliographystyle{ieeetr} 
\bibliography{citationv2}
\end{document}